\documentclass[
]{ceurart}

\usepackage{listings}
\usepackage{subcaption}
\begin{document}

\copyrightyear{2026}
\copyrightclause{Copyright for this paper by its authors.
  Use permitted under Creative Commons License Attribution 4.0
  International (CC BY 4.0).}

\conference{LIR'26: The 1st Late Interaction Workshop @ ECIR 2026, April 02, 2026, Delft, the Netherlands}

\title{A Brief Comparison of Training-Free Multi-Vector Sequence Compression Methods}


\address[1]{Johns Hopkins University, 3400 N. Charles St., Baltimore, MD 21218, USA}
\author[1]{Rohan Jha}[%
orcid=0009-0001-5008-4949,
email=rjha5@jh.edu,
url=https://www.cs.jhu.edu/~rjha5/,
]
\cormark[1]
\fnmark[1]

\author[1]{Chunsheng Zuo}[%
orcid=0009-0000-4988-2401,
email=czuo3@jh.edu,
url={https://scholar.google.com/citations?user=Uy79AH0AAAAJ},
]
\fnmark[1]

\author[1,2]{Reno Kriz}[
orcid=0000-0002-0239-9989,
email=rkriz1@jh.edu,
url=https://hltcoe.jhu.edu/researcher/reno-kriz/,
]
\address[2]{Human Language Technology Center of Excellence, 810 Wyman Park Drive,
Baltimore, Maryland 21211, USA}
\author[1]{Benjamin Van Durme}[
orcid=0000-0003-4328-4288,
email=vandurme@jhu.edu,
url=https://www.cs.jhu.edu/~vandurme/,
]

\cortext[1]{Corresponding author.}
\fntext[1]{These authors contributed equally.}


\begin{abstract}
While multi-vector retrieval models outperform single-vector models of comparable size in retrieval quality, their practicality is limited by substantially larger index sizes, driven by the additional sequence-length dimension in their document embeddings.
Because document embedding size dictates both memory overhead and query latency, compression is essential for deployment.
In this work, we present an evaluation of \textit{training-free} methods targeting the token sequence length, a dimension unique to multi-vector retrieval. 
Our findings suggest that token merging is strictly superior to token pruning for reducing index size while maintaining retrieval effectiveness.
\end{abstract}

\begin{keywords}
  Information Retrieval \sep
  Multi-Vector Retrieval \sep
  Index Compression \sep
  Token Pruning \sep 
  Token Pooling
\end{keywords}

\maketitle

\section{Introduction}
\label{sec:intro}

Multi-vector retrieval models, exemplified by ColBERT \cite{Colbert}, represent documents as sets of small token-level embeddings, enabling fine-grained semantic matching via late interaction. 
This approach consistently delivers stronger retrieval quality than single-vector models of comparable size, particularly in out-of-domain settings \cite{jha_2024_jinacolbertv2}. 
However, naively storing per-token embeddings for every document in a collection yields index sizes orders of magnitude larger than their single-vector counterparts, increasing both storage costs and query latency --- albeit not linearly in practice.

Compression is therefore essential to practical multi-vector deployment, and existing work has addressed this along several axes: reducing embedding precision, reducing embedding dimensionality, or reducing the number of embeddings stored per document. 
In practice, many effective methods operate across more than one of these axes --- for instance, product quantization jointly reduces dimensionality and precision (number of codebooks and code length), while PLAID \cite{santhanam_2022_plaid} combines centroid-based encoding with low-bit residuals to a similar effect. The third axis, token sequence length, is unique to multi-vector models and remains comparatively underexplored.
Methods along this axis fall into three broad families: fixed-length embeddings \cite{lassance_2021_StudyTokenPruning, macavaney_2025_constbert, xiao_2025_metaembed}, token pruning \cite{lassance_2021_StudyTokenPruning, acquavia_2023_StaticPruningMultiRepresentation, he_2025_tokenpruningoptimization}, which discards embeddings deemed uninformative, and token pooling \cite{hofstatter_2022_colberter, clavie_2024_reducing}, which merges groups of similar tokens into representative supervectors.

Crucially, several methods in the latter two families are training-free, making them attractive as drop-in post-hoc compression steps for any existing multi-vector model. 
Yet there is no direct comparison of these approaches under controlled conditions. 
In this work, we evaluate training-free token pruning and token merging methods across BEIR datasets \cite{thakur_2021_beir}, and confirm the intuition cited in \cite{clavie_2024_reducing} that \textbf{token pooling methods consistently outperform token pruning approaches across all compression ratios} in our examined text retrieval domains. Pooled sequence compression techniques often matched uncompressed performance with 2-5$\times$ fewer embeddings.

\section{Methods}
\label{sec:methods}

We study \emph{training-free}, document-side sequence compression for multi-vector retrieval.
A document is encoded as token embeddings $\mathbf{D} = [\mathbf{d}_1, \dots, \mathbf{d}_L]$; retrieval uses the MaxSim operator: $\operatorname{score}(\mathbf{Q}, \mathbf{D}) = \sum_{i} \max_{j}\; \mathbf{q}_i^\top \mathbf{d}_j$.
Compression replaces $\mathbf{D}$ with a shorter $\widetilde{\mathbf{D}}$ of length $\widetilde{L} < L$, reducing both index size and consequently query latency.
Following \cite{clavie_2024_reducing}, the first token (the document marker token \texttt{[D]}) is protected from compression.

\paragraph{Token Pruning}

Pruning assigns each non-protected token a scalar importance score and retains only the top-$C$ tokens, discarding the rest.
We evaluate three scoring functions:
\textbf{(i)~Random} selection as a stochastic baseline;
\textbf{(ii)~Attention} \cite{lassance_2021_StudyTokenPruning}, which scores each token by the total attention it receives across all heads and source positions in the final transformer layer; and
\textbf{(iii)~IDF} \cite{lassance_2021_StudyTokenPruning, acquavia_2023_StaticPruningMultiRepresentation}, scoring tokens by corpus-level inverse document frequency to preferentially retain rare terms.
Targeting a fixed compression ratio, IDF pruning is applied \textit{document-wise} using globally collected IDF statistics.

\paragraph{Token Pooling}

Pooling groups non-protected tokens into $C$ clusters and replaces each cluster with the mean of its member embeddings, preserving information rather than discarding it.
Clusters are formed by assigning each token to its nearest \emph{anchor}; we select the $k$ anchors via the top-$k$ scores according to \textbf{Random}, \textbf{Attention}, or \textbf{IDF} (as above).
We additionally evaluate two clustering-based variants that partition tokens directly:
\textbf{Spherical $k$-Means} on $\ell_2$-normalised embeddings, and
\textbf{Hierarchical (Ward) clustering} with cosine distance as proposed in \cite{clavie_2024_reducing}.

\paragraph{Experimental Setup}
We compare an uncompressed baseline against our compression methods across 5 keep ratios ($r \in \{0.10, 0.20, 0.33, 0.50, 0.75\}$) on six BEIR \cite{thakur_2021_beir} datasets (\texttt{arguana}, \texttt{scidocs}, \texttt{fiqa}, \texttt{nfcorpus}, \texttt{scifact}, \texttt{trec-covid}) and three CoIR \cite{li_2025_coir} text-to-code retrieval datasets (\texttt{apps}, \texttt{cosqa}, \texttt{synthetic-text2sql}), reporting NDCG@10 and Recall@100 via \texttt{ranx} \cite{bassani_2022_ranx}.
Our compression targets only document embeddings, which are encoded with a maximum length of $L=300$ tokens for BEIR, and $L=2048$ for CoIR datasets.
We use the default query lengths in \cite{chaffin_2025_pylate}'s BEIR baseline for BEIR datasets and 1024 query tokens for CoIR datasets.
Both settings' models use no \texttt{[MASK]} query expansion tokens \cite{khattab_2020_colbert} during encoding.
After compression, embeddings are $\ell_2$-normalized and indexed with PLAID.

\section{Results}
\label{sec:results}
\begin{figure*}[ht!]
    \centering
    \includegraphics[width=\textwidth]{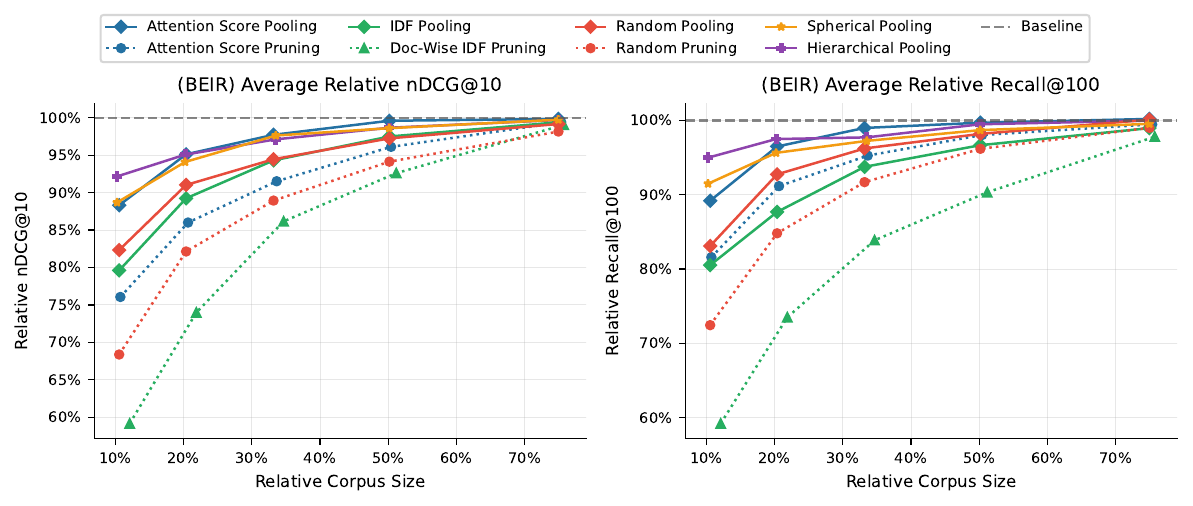}
    \caption{nDCG@10 (left) and Recall@100 (right) relative to Baseline (no compression) averaged across 6 BEIR datasets. Pooling methods (solid lines) consistently degrade much more gracefully than pruning methods (dotted lines) under heavy compression.}
    \label{fig:average_performance_combined}
\end{figure*}

\begin{figure*}[ht!]
    \centering
    \includegraphics[width=\textwidth]{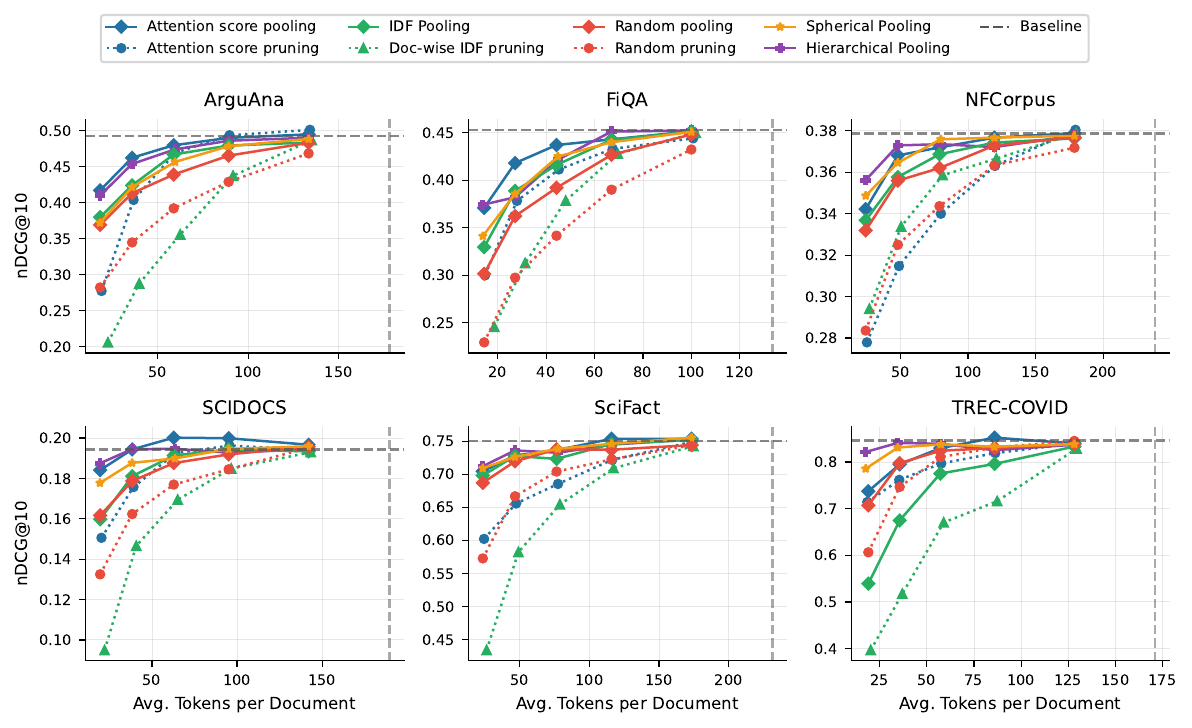}
    \caption{Dataset-specific breakdowns of nDCG@10 vs. average document length on 6 BEIR datasets. Axes are scaled individually to highlight per-dataset compression characteristics.}
    \label{fig:combined_ndcg}
\end{figure*}

\begin{figure*}[ht!]
    \centering
    \includegraphics[width=\textwidth]{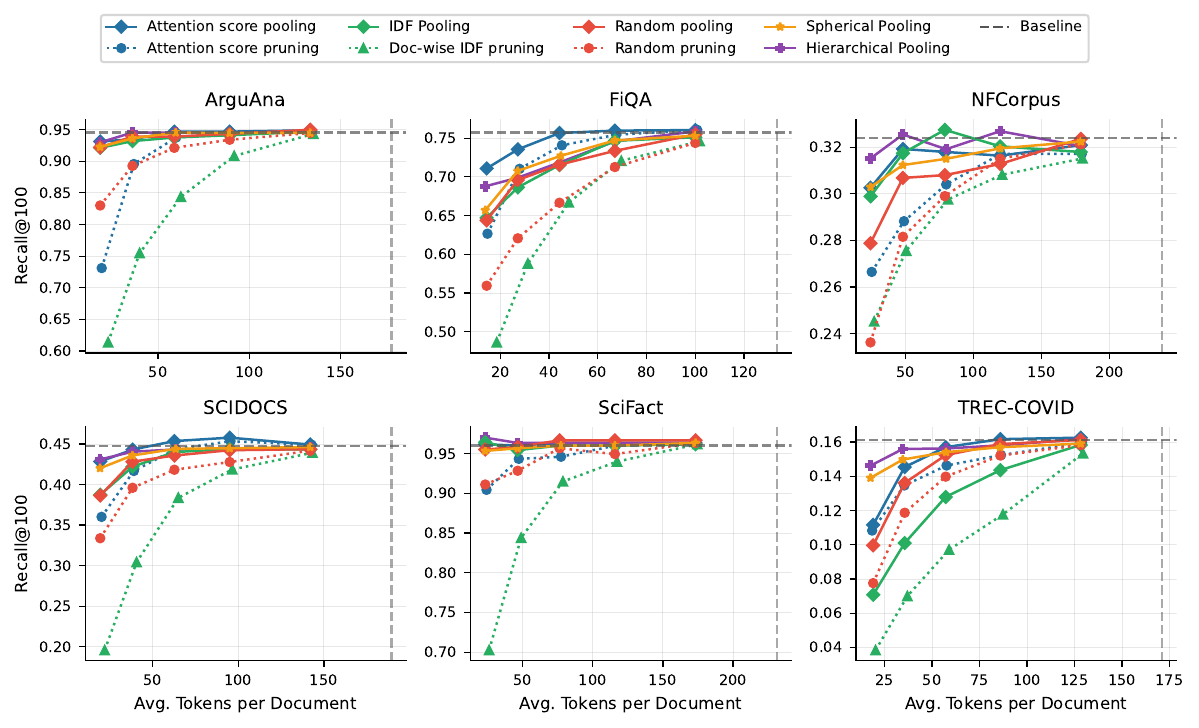}
    \caption{Dataset-specific breakdowns of Recall@100 vs. average document length on 6 BEIR datasets. Axes are scaled individually to highlight per-dataset compression characteristics.}
    \label{fig:combined_recall}
\end{figure*}

Our evaluation across the BEIR benchmark reveals a consistent and distinct advantage for token pooling (merging) over token pruning methods, particularly at aggressive compression rates. 
Furthermore, our results challenge several conventional assumptions regarding embedding importance and clustering complexity. 
The overall trends for nDCG@10 and Recall@100 across different keep ratios are presented in Figure~\ref{fig:average_performance_combined}, while dataset-specific breakdowns are provided in Figures~\ref{fig:combined_ndcg} and \ref{fig:combined_recall}.

The empirical results strongly support the hypothesis that aggregating semantic information is strictly superior to discarding it. Token pooling methods preserve the underlying semantic density of the original document embeddings much better than their pruning counterparts. 

Our results show that pooling can achieve higher compression ratios at equivalent retrieval quality. 
For instance, across the averaged BEIR datasets, Hierarchical Pooling at a keep ratio of $r=0.20$ (averaging roughly 38 tokens per document) achieves an average nDCG@10 of 0.497 (95.7\%) and Recall@100 of 0.588 (98.1\%). 

In contrast, IDF Pruning requires a much larger keep ratio of $r=0.50$ (95 tokens per document) just to reach an nDCG@10 of 0.474 (92.4\%) and a Recall@100 of 0.593 (98.9\%). 
In practical settings, pooling allows for more than double the index size reduction compared to pruning at the same target performance threshold.

At severe compression levels ($r \le 0.20$), the divergence between the two methods becomes catastrophic for pruning. Discarding 80\% to 90\% of a document's tokens removes critical exact-match and semantic cues necessary for the MaxSim operator to function effectively. Pooling methods, however, mitigate this by ensuring that even if a specific token is removed, its semantic footprint is partially retained within a cluster's pooled vector. Consequently, the performance floor for pooling methods remains significantly higher under extreme storage constraints.

\begin{figure*}[h!]
    \centering
    \includegraphics[width=\linewidth]{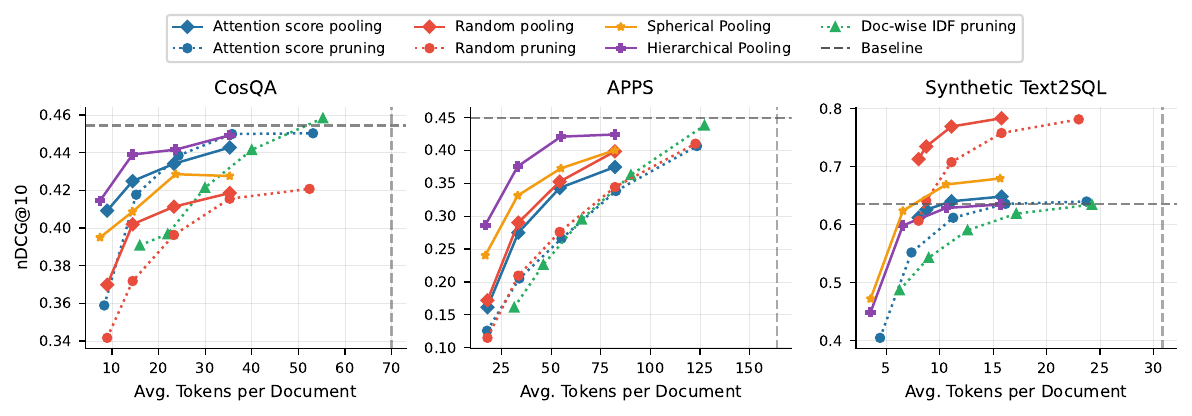}
    \caption{nDCG@10 vs. average document length on 3 CoIR text-to-code retrieval tasks.}
    \label{fig:coir}
\end{figure*}

Furthermore, while we confirm \cite{clavie_2024_reducing}'s finding of Hierarchical Pooling being the best overall compression method, our findings emphasize that the best method may vary with the task at hand.
For instance, in three of the tested BEIR datasets (\texttt{arguana}, \texttt{scidocs}, \texttt{scifact}) simple Attention Pooling achieves similar or better performance, while being cheaper to compute than $k$-Means or Hierarchical Pooling.
While the cost of offline indexing is generally considered less important in retrieval tasks --- as it does not directly impact query latency in large-scale web search --- this assumption breaks down in the setting where LLM agents begin to construct and update search indices ad hoc over arbitrary collections of data as an intermediate step in their tasks. 
In such settings, the cost of compression itself should be considered alongside the size of the compressed representation.\footnote{Attention Pooling: $\mathcal{O}(L \cdot \widetilde{L} \cdot d)$; Spherical $k$-Means ($I$ iterations): $\mathcal{O}(I \cdot L \cdot \widetilde{L} \cdot d)$; Hierarchical Pooling: $\mathcal{O}(L^2 \cdot d + L^2 \log L)$.}
Given that Hierarchical Pooling exhibits a greater gap in performance over the other pooling methods in the case of CoIR (Figure \ref{fig:coir}), and coding has so far been a primary agentic task, this tension might be felt acutely.

\section{Conclusion}
\label{sec:conclusion}

This work presents a controlled evaluation of eight training-free sequence compression methods for multi-vector retrieval across six zero-shot BEIR datasets and 3 CoIR text-to-code retrieval tasks.
Our empirical results confirm \cite{clavie_2024_reducing}'s finding of the strong performance of pooling-based index compression mechanisms for multi-vector retrieval, quantifying their superiority over pruning-based approaches.


Our study is limited to document-side, training-free compression on \textit{text-only} data.
We acknowledge recent works \cite{yan_2025_docpruner, yan_2026_prunethenmerge} that identify pruning as a more beneficial method, instead of or before pooling in the \textit{visual document} setting.
We hypothesize that the apparently conflicting conclusions in our and these works are a function of the differences in informational density and distribution in varying modalities.
Future work should explore adaptive compression strategies that dynamically vary the keep ratio $r$ based on a document's semantic density or investigate similar QPP-style dynamic compression on the query side. 
As multi-vector models continue to see wider deployment, implementing efficient, principled compression techniques will remain essential for balancing state-of-the-art effectiveness with real-world usability.

\bibliography{references}



\end{document}